\documentclass[a4paper]{jpconf}
\usepackage{graphicx}
\begin{document}
\title{Multiplicities and bulk thermodynamic quantities at 
 \mathversion{bold}{$\sqrt{s_{NN}}=130$}  GeV with SHARE}
\author{Giorgio Torrieri},
\address{Department of Physics, University of Arizona, Tucson, Arizona, 85721, USA, and\\
Department of Physics, McGill University, Montreal, QC H3A-2T8, Canada}
\ead{torrieri@hep.physics.mcgill.ca}
\author{Johann Rafelski}
\address{Department of Physics, University of Arizona, Tucson, Arizona, 85721, USA}

\begin{abstract}
We introduce the  Statistical Hadronization with Resonances 
(SHARE) suite of  programs and perform a study of particle  multiplicities 
as well as bulk thermodynamic quantities for the RHIC Au--Au reactions  at $\sqrt{s_{NN}}=130$ GeV.
We also show that the statistical hadronization model, with parameters 
fitted to match pion, proton and hyperon ratios, in turn correctly
and consistently reproduces rapidity  particle multiplicity. 
\end{abstract}.

\section{Introduction}
The statistical hadronization model  \cite{Fer50,Pom51} has been used 
extensively to study soft strongly interacting particle production since the 1950s.
When the full spectrum of strongly interacting 
resonances is included  \cite{Hag65}, this approach is capable to 
 describe the abundances and spectra of the produced particles in detail.
The emitted particles carry information about 
 the gross features of the hadron source.  Their study allows
precise extrapolation to unmeasured particles and/or kinematic domains, allowing
 understanding of bulk properties of hadronizing matter such as 
particle multiplicity, mean energy per particle, specific per baryon
entropy etc. Once hadron spectra are understood, information about
the dynamical properties such as collective flow of hadronizing 
matter becomes accessible. 

 Other important physical information 
is contained in  the  statistical model parameters obtained  fitting 
measured particle ratios. These offer additional insights, for example
one may want to relate the chemical (particle production) freeze-out
temperature to the phase transformation temperature of the deconfined
quark--gluon plasma (QGP) phase into hadrons. The chemical freeze-out temperature also 
helps understand  the percentage of total particles
which originated from resonance decays. The chemical freeze-out parameters 
fix e.g. the charged and neutral hadronic multiplicities,  flavor density,
baryon stopping, and so on.

\begin{figure*}[!bt]
%\hspace*{3cm}
\hspace*{-.15cm}
\includegraphics[width=18pc,height=16pc]{gamma_s_plot.eps}
%\hspace*{.45cm}
\includegraphics[width=17pc,height=16.3pc]{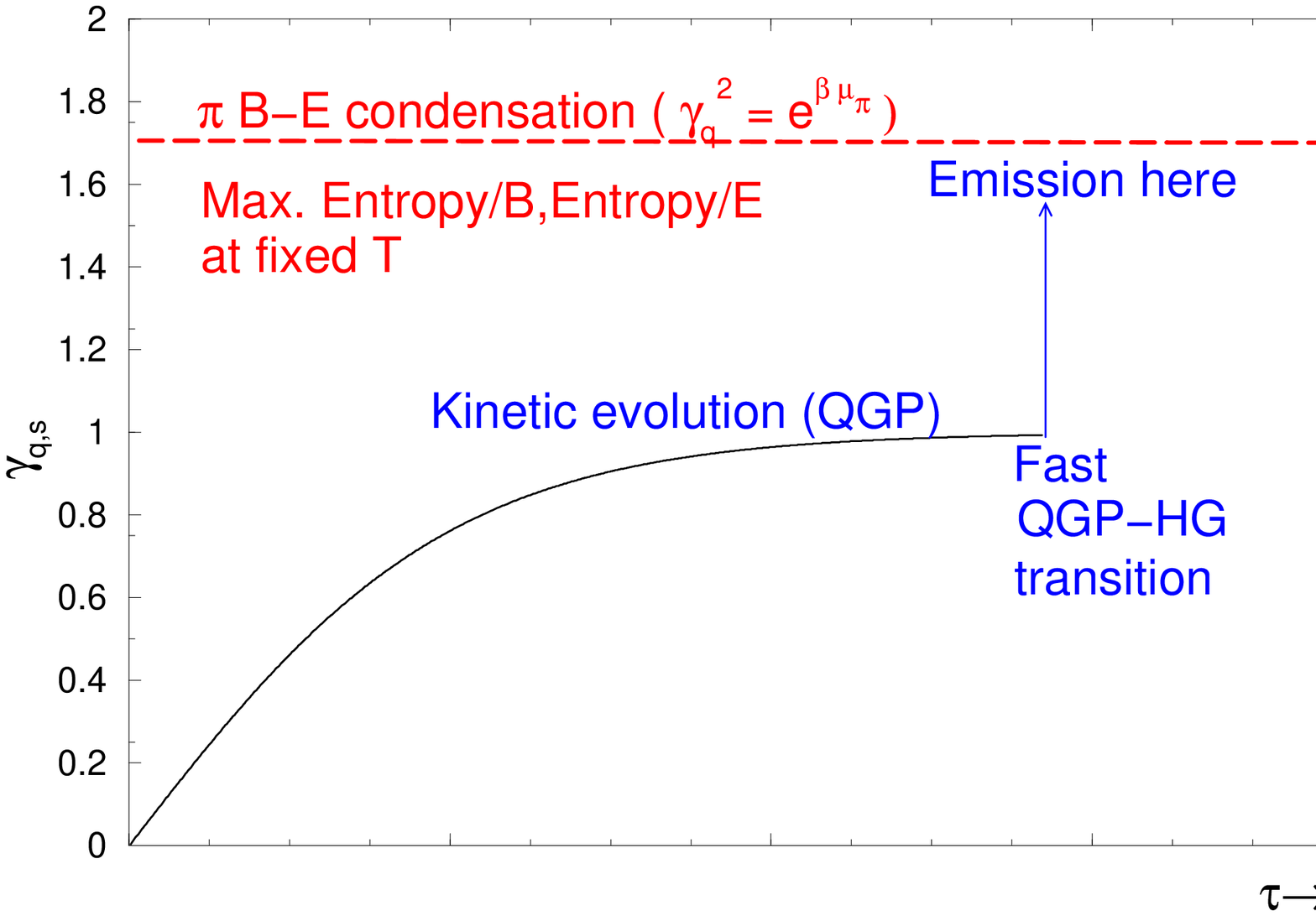}
\caption{\label{gammaqs}Chemical non-equilibrium can arise out
of kinetic evolution (left), and in rapid QGP  freeze-out into hadrons (right).}
\end{figure*}

To properly address these issues, standardization
 of the technical and mathematical 
tools  employed in statistical hadronization studies  has to occur.
 Furthermore the model  differences have to be  understood and the
different versions  have to be unified. Here,  we note the chemical 
equilibrium version, where all (light and strange) flavor
yields are assumed to have evolved for a long time in the 
hadron phase reaching equilibrium yield 
\cite{barannikova,florkowski,BDMRHIC}.
More refined approaches 
 allow for an under-saturation of the strangeness quantum number \cite{jan_gammas}, 
implemented quantitatively by a parameter $\gamma_s$ (assumed to be $<1$) which 
affects both $s$ and $\overline{s}$ in the same way
($\gamma_{s}=\gamma_{\overline{s}}$, in contrast to the fugacity 
$\lambda_{s}=\lambda_{\overline{s}}^{-1}$).
The simplest physical scenario applicable here is that the hadron source 
 does not live for long enough for strangeness production  to approach
 chemical equilibration (Fig. \ref{gammaqs} left), while the more rapidly
evolving light flavor yields had time to equilibrate.

The third model approach is to allow for
chemical non-equilibrium also in the light flavor yield. 
This situation is most likely to be found 
in a reaction scenario involving formation of the QGP.
Namely, the phase space density of flavors in  QGP is higher than  of a equilibrated 
hadron gas in the physical domain explored today at SPS and at RHIC \cite{jan_gammaq}. 
This is illustrated in Fig. \ref{gammaqs} in the right panel, where
we show that the  phase space density difference 
compensation leads to a step-up in the
quark flavor occupancy parameter $\gamma$.  

%%%%%%%%%%%%%%%%%%%%%%%%%%Figure 3
%%%%%%%%%%%%%%%%%%%%%%%%%%%%%%%%%%%%%

Linked to the question of chemical equilibration is the timescale between
hadronization (the moment at which degrees of freedom become hadronic) and
freeze-out (the moment at which hadrons stop interacting). It is generally assumed
that there are two freeze-outs, chemical and thermal, the former corresponding
to particle production and latter to spectral shape generated by  
elastic hadron--hadron scattering.  Chemical equilibration requires 
slow evolution of the hadronic gas system, hence to a situation 
 in which hadronization and freeze-out are
well separated, and considerable modification of all hadronic yields 
could occur in the interacting hadron gas phase \cite{barannikova}.
In such a scenario one also expects that the thermal freeze-out 
is clearly different from the chemical freeze-out.

If QGP hadronizes, 
the difference between quark and hadron phase spaces makes
 it likely that a sudden freeze-out will occur creating initially
hadrons  in chemical non-equilibrium;  In particular, the high 
entropy density of a QGP phase with massless degrees of freedom 
accompanied by the rapid matter flow driven by the high
internal pressure
makes it reasonable that  hadrons are produced
in a fast phase transformation, which leads to 
flavor  over-saturated state \cite{jan_gammaq},
 with more $q \overline{q}$ and $s \overline{s}$ pairs than 
is expected at chemical equilibrium.
It is necessary to ascertain phenomenologically
whether, in fact, one needs non-equilibrium to account 
for particle abundances in heavy ion collisions.
The Statistical Hadronization with Resonances (SHARE) 
package has been developed allowing to resolve this ambiguity
as one of the tasks

In the following section we give introduce the statistical 
model, and the details  implemented in SHARE.   We refer the user to
Ref. \cite{share} for a detailed presentation and user's manual.
This is followed in section \ref{results} by discussion
 of our results, addressing in turn
particle ratios and bulk properties. 

%%%%%%%%%%%%%%%%%%%%%%%%%%%%%%%%%%%%%%%%%%%%%%%%%%%%%%%%
\section{Main SHARE features}
Many of our experimental 
friends present at this meeting can obtain on back of an envelope the 
statistical yield of  stable hadrons. SHARE extends this
to include important refinements capable to change these
results significantly.

%%%%%%%%%%%%%%%%%%%%%
\subsection{Particle yields allow quantum statistics and chemical non-equilibrium}
The Grand-Canonical statistical prescription assumes 
that enough particles of each flavor are produced to keep
fluctuations of each quantum number small, and the system 
volume reduces to a ``large'' normalization constant.
This requirement, together with entropy maximization, 
leads to the Fermi--Dirac or Bose--Einstein
distribution functions for  densities of particle species~$i$:
\begin{eqnarray}
\label{nmdef}
n(m_i,g_i;T,\Upsilon_i)\equiv n_i &=& g_i \int {d^3p \over (2 \pi)^3} 
{1 \over \Upsilon^{-1}_i 
\exp(\sqrt{p^2+m^2_i} / T ) \pm 1} ,\\
&=& \frac{g_i}{2 \pi^2}  \sum_{n=1}^{\infty} (\mp)^{n-1} 
\Upsilon_i^{n} \, \frac{T \, m_i^2}{n} \,
K_2 \left( \frac{n m_i}{T} \right). 
\label{ni}
\end{eqnarray}
In Eq.\,(\ref{ni}), the upper
signs refer to fermions and the lower signs to bosons, respectively.  
$\Upsilon_i$ is the fugacity factor, and $m_i$ is the particle mass. The 
quantity $g_i=(2J_i+1)$ is  the spin degeneracy factor as we distinguish 
all particles according to their electrical charge and mass.  The index $i$ 
labels different particle species, including hadrons which are stable under 
strong interactions (such as pions, kaons, nucleons or hyperons) and hadron 
which are unstable ($\rho$ mesons, $\Delta(1232)$, etc.). 
The second form, Eq.\,(\ref{ni}), expresses the momentum integrals in
terms of the modified Bessel function $K_2$. This form is practical in
the numerical calculations and is used in the SHARE code. Although
in principle in some limiting cases this is not a convergent expansion,
this is a  rare exception:  the 
series expansion (sum over $n$) converges when $\Upsilon_ie^{-m_i/T}<1$.
Violation of this condition occurs in practical context only for the pion case 
within the range of parameters of interest. 

In the most general chemical condition\footnote{This condition is commonly called  chemical 
non-equilibrium. However,  the conventional equilibrium 
in which existent  particles are redistributed according to chemical 
potentials is maintained here. The  non-equilibrium regarding particle 
production is, in precise terms,  called absolute chemical 
(non)equilibrium.}, the fugacity is defined through the parameters
$\lambda_{I^{\,i}_3}, \lambda_q, \lambda_s, \lambda_c$ (expressing, 
respectively, the isospin, light, strange and charm quark fugacity factors), 
and $\gamma_q,\gamma_s,\gamma_c$ (expressing 
the light, strange and charm quark phase space occupancies, $=1$ for 
absolute yield equilibrium).
The fugacity $\Upsilon_i$ is then  given by:
\begin{equation}
\Upsilon_i=\lambda_{I^i_3} \left(\lambda_q \gamma_q\right)^{N^i_q}
\left(\lambda_s \gamma_s\right)^{N^i_s} 
\left(\lambda_{c} \gamma_{c}\right)^{N^i_{c}}
\left(\lambda_{\bar q} \gamma_{\bar q}\right)^{N^i_{\bar q}}
\left(\lambda_{\bar s} \gamma_{\bar s}\right)^{N^i_{\bar s}}
\left(\lambda_{\bar c} \gamma_{\bar c}\right)^{N^i_{\bar c}}, 
\label{upsilons}
\end{equation}
where
\begin{equation}
\lambda_{q}=\lambda^{-1}_{\bar q},\qquad
\lambda_{s}=\lambda^{-1}_{\bar s},\qquad 
\lambda_{c}=\lambda^{-1}_{\bar c},\qquad 
\label{lambdas}
\end{equation}
and
\begin{equation}
\gamma_{q}=\gamma_{\bar q},\qquad
\gamma_{s}=\gamma_{\bar s},\qquad
\gamma_{c}=\gamma_{\bar c}.
\label{gammas}
\end{equation}
Here, $N^i_q$, $N^i_s$ and $N^i_c$ are the numbers of light $(u,d)$, strange
$(s)$  and charm $(c)$ quarks in the $i$th hadron, and $N^i_{\bar q}$, $N^i_{\bar
s}$  and $N^i_{\bar c}$ are the numbers of the corresponding antiquarks 
in the same hadron.

%%%%%%%%%%%%%%%%%%%%%%%%%%%%%%%%%%%%%%%%%%%%%%%%
\subsection{Particle yields from resonance decays}
\label{resdec}
At first, we consider hadronic resonances as if they 
were particles with a given well defined mass, {\it e.g.,} their decay width
is insignificant. All hadronic resonances decay rapidly after freeze-out, 
feeding the stable particle abundances.  Moreover,  heavy resonances 
may decay in cascades, which are
implemented in the algorithm where all decays proceed sequentially
from the heaviest to lightest particles. As a consequence, the light
particles obtain contributions from the heavier particles, which have
the form 
\begin{eqnarray}
n_1  &=& b_{2\rightarrow 1} 
 \, ... \, b_{N\rightarrow N-1}
n_N,
\label{n1}
\end{eqnarray}
where $b_{k\rightarrow k-1}$ combines the branching ratio for the $k
\rightarrow k-1$ decay (appearing in \cite{pdg}) with the appropriate
Clebsch--Gordan coefficient.  The latter accounts for the isospin symmetry
in strong decays and allows us to treat separately different charged
states of isospin multiplets of particles such as 
nucleons, Deltas,  pions, kaons, {\it etc}. For
example, different isospin multiplet member states of 
$\Delta$ decay according to the following
pattern:
\begin{eqnarray}
&& \Delta^{++} \rightarrow \pi^+ + p, 
\label{D++} \\
&& \Delta^{+} \rightarrow  {1 \over 3} (\pi^+ + n) + {2 \over 3}
(\pi^0 + p), 
\label{D+} \\
&& \Delta^{0} \rightarrow  {1 \over 3} (\pi^- + p) + {2 \over 3}
(\pi^0 + n), 
\label{D0} \\
&& \Delta^{-} \rightarrow  \pi^- + n.  
\label{D-}
\end{eqnarray}
Here, the branching ratio is 1 but the Clebsch--Gordan coefficients
introduce another factor leading to the effective branching ratios
of 1/3 or 2/3, where appropriate.

To implement this procedure on every particle, one needs to keep in mind that
the partial widths 
(product of branching ratio with total width) are often not
sufficiently well known.     In addition, in case of weak decays, an
experiment-specific acceptance coefficient is needed to correctly model
the observed particle rate. This introduces implementation dependent 
variances of statistical hadronization.  To combat invisible 
model variations we  suggest  an open-source ``standard'', 
where resonance decay trees are kept on record and can be  updated
in a transparent fashion, and weak acceptances can be
set to correspond to needs of each experiment.  This is 
the SHARE code \cite{share}.      

In SHARE, as a rule, all decays with the branching ratios smaller 
than 1\% are disregarded. In addition, if the
decay channels are classified  as {\it dominant}, {\it large, seen}, or
{\it possibly seen,}  the most important
channel is taken into account. If two or more channels 
are said to be equally important,
we take all of them with the same weight. For example $f_{0}(980)$
decays into $\pi \pi $ (according to \cite{pdg} this is the {\it
dominant} channel) and $K \overline{K}$ (according to \cite{pdg} this
is the {\it seen} channel). In our approach, according to the rules
stated above, we include only the process $f_{0}(980) \longrightarrow
\pi \pi $. Similarly, $a_{0}(1450)$ has three decay channels: $\eta
\pi $ ({\it seen}), $\pi \eta ^{\prime }(958)$ ({\it seen}), and $K
\overline{K}$ (again {\it seen}). In this case, we include all three
decay channels with the weight (branching ratio)~1/3.
A table of allowed decays and branching ratios is provided and 
can be updated. Users can modify this table to study the magnitude
of systematic error introduced by incomplete knowledge of both
resonance masses and decay parameters.

%%%%%%%%%%%%%%%%%%%%%%%%%%%%%%%%%%%%%%%%%%%%%%%%%%%%
\subsection{Hadron  yields  allowing for  finite resonance width}
\label{widthdis}

If the particle $i$ has a finite width $\Gamma_i$, the thermal yield of
the particle is more appropriately obtained by weighting Eq.\,(\ref{nmdef})
over a range of masses to take the mass spread into account:
\begin{equation}
\tilde n_i^{\Gamma}=\int\! dM\, 
n(M,g_i;T,\Upsilon_i)\frac{1}{2\pi}\frac{\Gamma_i}{(M-m_i)^2+\Gamma_i^2/4}
\to n_i, \mbox{\ \ for\ \ } \Gamma_i\to 0.
\label{widthn1}
\end{equation}
The use of the  Breit--Wigner  distribution with energy independent width means 
that there is a finite probability that the resonance would be formed at 
unrealistically small masses. Since the weight involves a thermal distribution 
$n(M,g_i;T,\Upsilon_i)$ which would contribute in this unphysical domain, one 
{\it must} use, in Eq.\,(\ref{widthn1}), an energy dependent width. 

The dominant  energy dependence of the width is due to the decay threshold energy 
phase space factor, dependent  on the angular momentum present  in the decay. 
The explicit form can be seen in the corresponding reverse production 
cross sections \cite{phase1,phase2}. The energy dependent partial width in the 
channel $i\to j$  is to a good approximation: 
\begin{equation}
\label{partwid1}
\Gamma_{i\to j}(M)=b_{i\to j}\Gamma_i
     \left[1-\left(\frac{m_{ij}}{M}\right)^2\right]^{l_{ij}+\frac{1}{2}},
          \quad \mbox{for} \quad M>m_{ij}.
\end{equation}
Here, $m_{ij}$ is the threshold of the decay reaction with branching 
ratio $b_{i\to j}$. For example  for the decay 
of $i:=\Delta^{++}$ into $j:=p+\pi^+$, we have $m_{ij}=m_p+m_{\pi^+}$, while the
branching ratio is unity and the angular momentum released in 
decay is $l_{ij}=1$. From these partial widths 
the total energy dependent width arises,
\begin{equation}
\Gamma_i\to \Gamma_i(M) = \sum_{j}\Gamma_{i\to j}(M).
\end{equation}
For a resonance with width, we thus have replacing Eq.\,(\ref{widthn1}):
\begin{equation}
 n_i^{\Gamma} = \frac{1}{N_i} \sum_{j} 
\int_{m_{ij}}^{\infty} dM \, 
n(M,g_i;T,\Upsilon_i)\frac{\Gamma_{i\to j}(M)}{(M-m_i)^2+[\Gamma_i(M)]^2/4},
\label{widthn}
\end{equation}
and the factor $N$ (replacing $2\pi$) ensures the normalization:
\begin{equation}
\label{norm}
N_i= \sum_{j} 
\int_{m_{ij}}^{\infty}  dM 
\frac{\Gamma_{i\to j}(M)}{(M-m_i)^2+[\Gamma_i(M)]^2/4}.
\end{equation}

In principle, Eq.\,(\ref{widthn}) does not take into account the possibility
that the state into which one is decaying is itself a unstable state in 
a thermal bath. Doing this would require a further average over the width 
distribution of receiving state. This higher order effect is at present 
not implemented in SHARE. 

%%%%%%%%%%%%%%%%%%%%%%%%%%%%%%%%%%%%%%%%%%%%%%%%%%%%%%%%%%%%%%%%%%%%%%%
\subsection{Canonical effects}
 We did not address in SHARE refinement  specific
to small particle numbers where the grand canonical phase space
description fails \cite{canon}. 
Thus SHARE is geared to describe particle multiplicities
where within the causal interaction domain there is an effective 
`particle' heat bath. To be more specific, the grand canonical phase 
space will yield correctly e.g. the $\overline{\Omega}$ yield  in 
the limit that the number of strange quark pairs significantly exceeds
those required to make this particle, e.g. we need about 10 $s\bar s$ 
pairs. Without doubt these are the conditions prevailing in the 
physical environment we are exploring here, since at RHIC 130GeV  about
8 strange quark pairs are produced per each participating 
baryon \cite{firstRHIC,kpi1},
and baryon density is about 25 per unit of rapidity.  

%%%%%%%%%%%%%%%%%%%%%%%%%%%%%%%%%%%%%%%%%%%%%%%%%%%%%%%%%%%%%%%%%%%%%%%
\section{RHIC 130 fit results}\label{results}
%%%%%%%%%%%%%%%%%%%%%%%%%%%%%%%%%%%%%%%%%%%%%%%%
\subsection{Particle ratios}\label{ratios}
We study particle ratios  taken at different though
similar   collision centralities, e,g, 5\%, 6\%, 8\%, 10\%. In first and hopefully 
good approximation we expect
that the physical conditions established are similar in all these case. In other 
words we expect that for particle yields in the  statistical model to be consistent:
\begin{itemize}
\item  The volume should appear as a normalization constant.   Temperature and the other thermal parameters should be, within error, independent of the centrality for a range of
centralities.
\item  This normalization constant should present an approximately 
proportional dependence on the total charge multiplicity.
\end{itemize}

\begin{figure*}[!bt]
%\hspace*{3cm}
\hspace*{-.15cm}
\includegraphics[width=18pc,height=15pc]{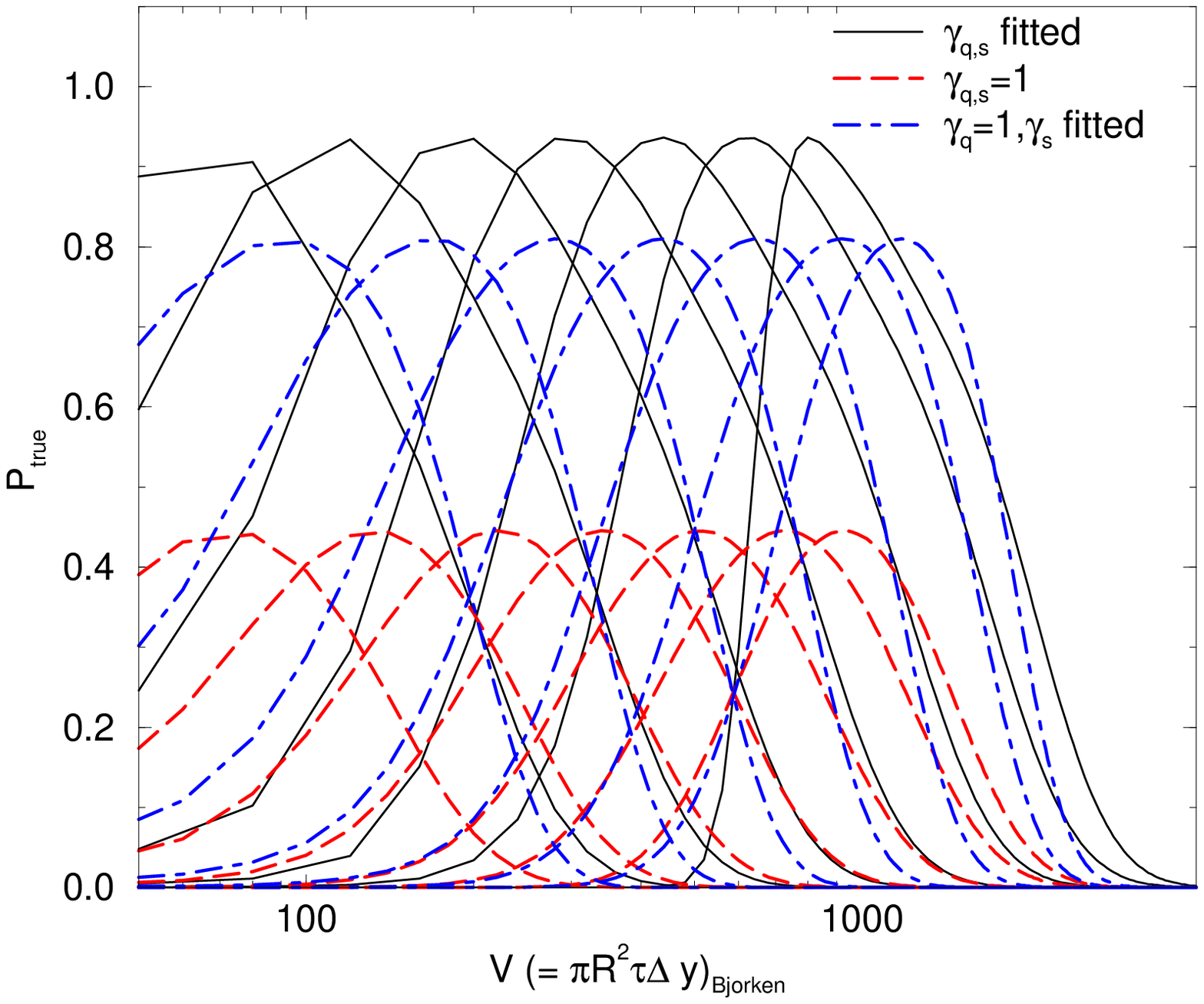}\hspace*{.2cm}
\includegraphics[width=18pc,height=14.5pc]{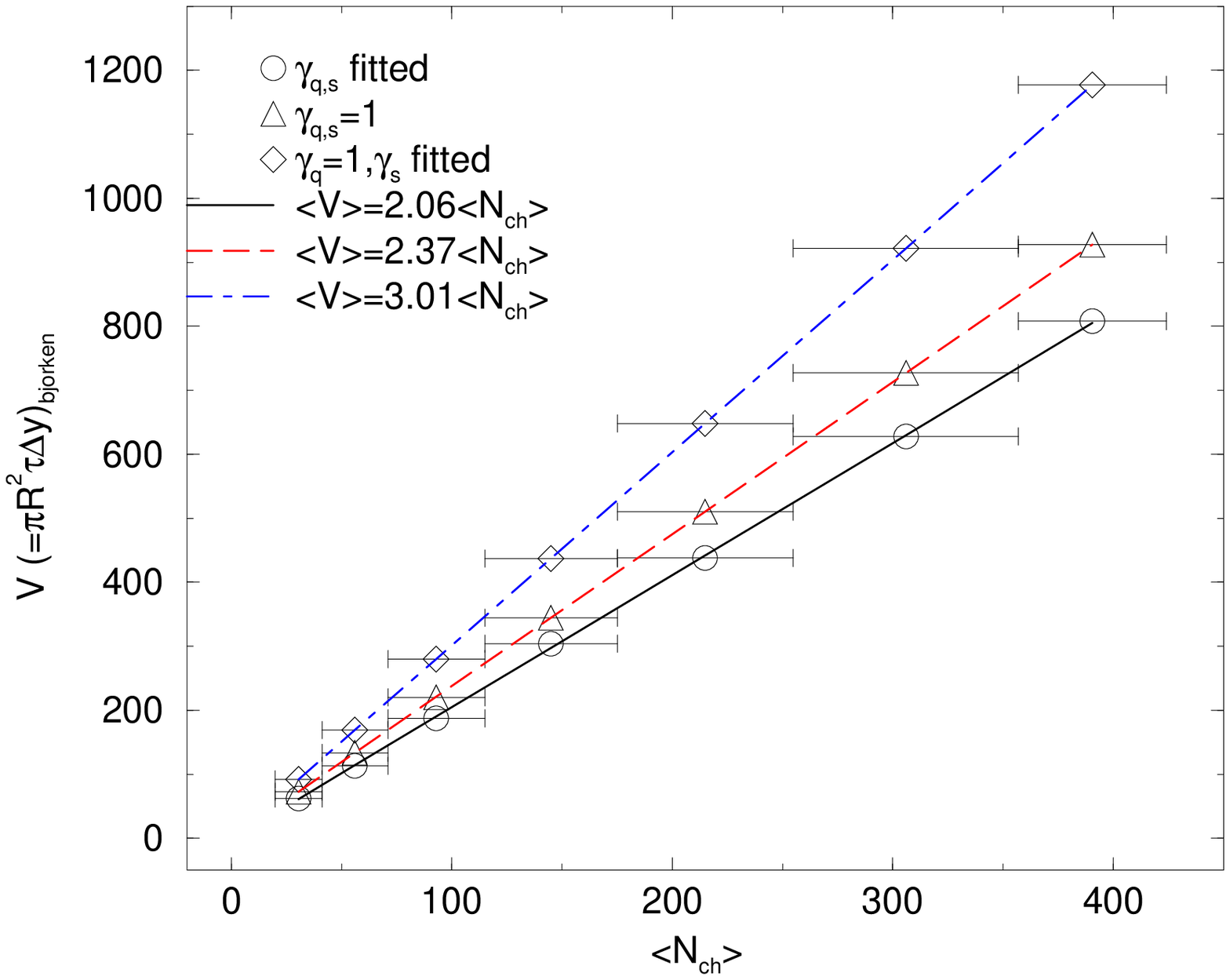}
\caption{\label{plot_nr} 
Statistical significance profiles for volume for different centralities 
(left), and it's correlation with the charged
particle number (right).}
\end{figure*}

\begin{figure*}[!bt]
\begin{center}
%\hspace*{3cm}
\hspace*{-.15cm}
\includegraphics[width=36pc,height=20pc]{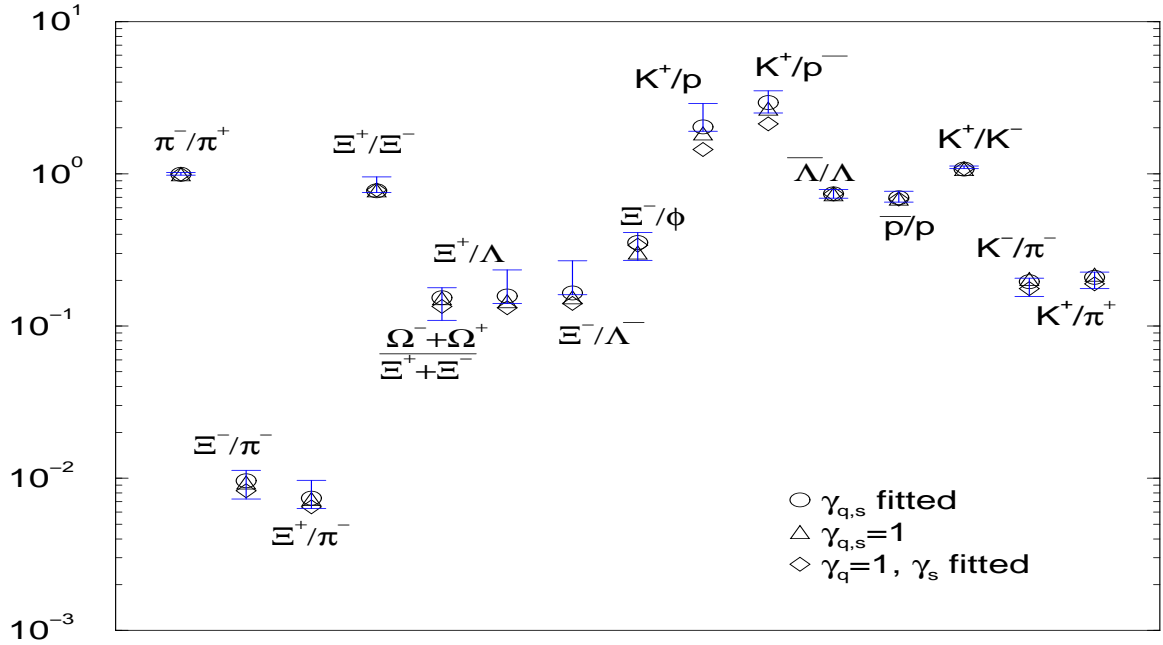}
\caption{\label{bestfit} 
The best fits to hadron yields  at RHIC-130 GeV. See table \ref{results_table130} for 
parameters of the fits and statistical significance.}
\end{center}
\end{figure*}

These requirements can be tested by fitting
the total particle multiplicity of a range of experimentally measured centrality bins, together with
a set of hadron ratios (Fig \ref{bestfit}).
We consider  seven most central bins, as measured by  STAR, \cite{starmult}, 
and find  no  significant variation of the fitted bulk parameters ($T,\lambda_{q,s},\gamma_{q,s},\lambda_{I3}$) and their errors.
The only parameter in the fit which does vary from bin to bin 
is the absolute normalization, needed to describe the charged
multiplicities;  It's statistical significance
profiles are shown in Fig. \ref{plot_nr} (left);  
 As noted elsewhere \cite{jan_gammaq,future}, 
introducing $\gamma_q$ significantly and consistently 
improves the statistical significance of the fit.

%%%%%%%%%%%%%%%%%%%%%%%%%%Figure 1
%%%%%%%%%%%%%%%%%%%%%%%%%%%%%%%%%%%%%n

When the fitted volume normalization parameter
 is plotted against the mean charged particle multiplicity, 
an approximately linear dependence is
observed for all models (Fig.~\ref{plot_nr} right).
Light quark non-equilibrium decreases the necessary volume by as much as 30 $\%$.
While this is predictable, since the non-equilibrium model invariably
 leads to quark over-saturation ($\gamma_q>1$) at high energies and
 thus higher particle density  \cite{jan_gammaq}. The smaller volume 
required in non-equilibrium has not as yet been subject to an in-depth 
analysis, particularly in light of the HBT puzzle \cite{hbt1}:
It is sometimes thought that an explosive hadronization scenario 
is ruled out by HBT measurements \cite{hbt2}. However,  as this study 
the   non-equilibrium particle production is  yielding a higher
 density than that predicted by equilibrium hydrodynamics models, 
which should  rectify this discrepancy.

We have used RHIC ratios (130 GeV) to calculate thermal parameters within each model.
While a plot, showing a model-data comparison, is given in Fig. \ref{bestfit} and table \ref{results_table130},  we refer the reader to \cite{future}
for an in-depth data and statistical significance analysis of the fits.

%%%%%%%%%%%%%%%%%%%%%%%%%%Figure 3

%%%%%%%%%%%%%%%%%%%%%%%%%%%%%%%%%%%%%
\begin{table*}[bt]
\caption{
\label{results_table130} RHIC-130 GeV hadronization parameters}
\vspace*{0.2cm}\small
\begin{tabular}{|c| c c| c c| c c|}
\hline
 &        \multicolumn{2}{c}{$\gamma_{q,s}$ vary }  &\multicolumn{2}{c}{$\gamma_{q,s}=1$ }  & \multicolumn{2}{c|}{$\gamma_{q}=1$, $\gamma_s$ varies }  \\
parameter & $\Gamma=0$ & $\Gamma$  \cite{pdg} & $\Gamma=0$ & $\Gamma$  \cite{pdg} & $\Gamma=0$ & $\Gamma$  \cite{pdg}\\
\hline
T [MeV]         & 133 $\pm$ 10 &  135 $\pm$ 12 & 158 $\pm$ 13 & 0.157 $\pm$ 15 & 152 $\pm$ 16 & 153 $\pm$ 23 \\
$10^4 (\lambda_{q}-1)$     & 708 $\pm$ 342 & 703 $\pm$  337  &   735 $\pm$ 390 & 730 $\pm$  382 & 724 $\pm$  373 &  721 $\pm$     363\\
$\lambda_s$\footnote{Found by solving for strangeness}  &    1.03132973   & 1.03203555 &  1.02636207 & 1.02788897 &1.0295346 &  1.0300848  \\
$\gamma_{q}$    & 1.66 $\pm$ 0.013 &  1.65 $\pm$ 0.030  & 1 & 1 & 1 & 1 \\
$\gamma_{s}$      & 2.41 $\pm$ 0.61 &   2.28 $\pm$  0.46  & 1 & 1 & 1.17 $\pm$ 0.30 &1.10  $\pm$ 0.25\\ 
$10^4 (\lambda_{I_3}-1)$ &  30 $\pm$ 305 & 28 $\pm$  293 & 59 $\pm$  564 & 53 $\pm$  508 &  64  $\pm$  525 &  59  $\pm$    481 \\
\hline
 & \multicolumn{6}{c|}{fit relevance}\\ 
\hline 
$N-p=\mbox{DoF}$                                               & 16-5 &    16-5   & 16-3   & 16-3   & 16-4 & 16-4 \\   
 $\chi^2/$DoF             &  0.4243  & 0.4554 &  1.0255   & 0.8832 &  0.6067 &  0.7301 \\ 
 significance   & 0.9461   & 0.9307 &  0.4225  &  0.5705 &  0.8385 &  0.7232 \\
\hline
\end{tabular}
 \end{table*}

%%%%%%%%%%%%%%%%%%%%%%%%%%%%%%%%%%%%%%%%%%%%%%%%%%%%%%%%%%%%%%%%%%%%%%%%%%%%%%%%%%%%%%%%%%
\subsection{Bulk quantities}\label{bulk}

We proceed to calculate $\sqrt{s_{NN}}=130$  GeV Au--Au bulk quantities.
The results are shown in table \ref{ext_table130}.
It is immediately apparent that many of the observed 
quantities depend strongly on whether light quarks are in
equilibrium. 
By contrast, introducing incomplete strange chemical 
equilibration while maintaining $\gamma_q =1$ does not produce
a significant shift in the calculated quantities from
 the full equilibrium values. This is due to the fact that 
in this case the fit leads to practically fully equilibrated 
system as within error $\gamma_s=1$ results.

Overall, the greatest change is that the entropy 
density is considerably higher within the $\gamma_q \ne 1$ model.
This is consistent with the model's physical motivation, 
since the high $\gamma_q$ is invoked to conserve entropy
in a QGP$\rightarrow$ HG transition, without need for
a mixed phase and expansion.

Strangeness per entropy, being primarily  sensitive to 
input ratios such as $K/\pi$ (fitted), 
is not significantly affected by model choice 
\cite{kpi1,kpi}.
Quantities such as particle multiplicity per baryon 
(which follows entropy per baryon), net charge per baryon, and strangeness
per baryon, however, vary considerably between models, making them promising
probes for statistical model variants involving chemical (non)equilibration.    
We have also show  (last line in table \ref{ext_table130} ) 
that the fraction of particles coming from weak decays
 varies considerably from
model to model, decreasing as non-equilibrium is introduced.      
 This is to be expected, since all non-equilibrium
fits yield over-saturated phase space occupancies.    
These enhance all particles, including resonances, thereby
increasing the percentage of particles emitted in strong decays. 
This feature should  also provide a  test for non-equilibrium: 
 Since all weak decays, including those which can not be 
reconstructed, tend to occur at a measurably large
distance from the primary vertex, a precise estimate 
of the number of particles which do not come from the primary
vertex could also serve as a potential probe for equilibrated emission.

Due to the higher particle multiplicity in the non-equilibrium case, 
the thermalized energy per particle goes down.
While this model does not include collective (transverse and longitudinal) 
flow, and hence can not address the total energy in the system (some of which
is contained in the collective motion, rather than the internal energy), the
discrepancy in the energy per particle can become a strong constraint 
if the models under consideration are coupled to a 
 model (such as the commonly used ``Blast Wave'') incorporating flow.
Successful fits of abundances and spectra within the same model
 have been made both in the equilibrium and
non equilibrium cases \cite{torrieri_njp,florkowski,comparison},
 and it will be interesting to rigorously test those for
full energy conservation.

%%%%%%%%%%%%%%%%%%%%%%%%%%%%%%%%%%%%%%%%%%%%%%%%%%%%%%%%%%%%%%%%%%%%%%%%%%%%%%%%%%%%%%%%%%
\begin{table*}[bt]
\caption{
\label{ext_table130} RHIC 130 GeV Au--Au best fit bulk quantities ratios, including charged particles ($h^{\pm 0}$,energy ($E$), Baryon/antibaryon ($B,\overline{B}$), strangeness ($s,\overline{s}$) and entropy (S).
The bottom ratio refers to the fraction of totally emitted particles which arise out of weak decays
(assuming negligible $K_L \rightarrow 3 \pi$ detectability)
}\vspace*{0.2cm}%\small
\begin{tabular}{|l| c c| c c| c c|}
\hline
 &      \multicolumn{2}{c}{$\gamma_{q,s}$ vary }  &\multicolumn{2}{c}{$\gamma_{q,s}=1$ }  & \multicolumn{2}{c|}{$\gamma_{q}=1$, $\gamma_s$ varies }  \\
ratio  & $\Gamma=0$ & $\Gamma$  \cite{pdg} & $\Gamma=0$ & $\Gamma$  \cite{pdg} & $\Gamma=0$ & $\Gamma$  \cite{pdg}\\
\hline
$(h^++h^-)/(B-\overline{B})$                       & 30.0827   & 30.821  &18.613 & 20.087&22.149  & 22.758   \\
$(h^+-h^-)/(B-\overline{B})$                       & 0.618   & 0.618  &0.603  & 0.602 &0.632  &  0.624     \\
$E/(B-\overline{B})$                               & 37.220   & 38.152 &31.756  & 33.403&36.004 & 36.441  \\    
$(s+\overline{s})/(B-\overline{B})$                &9.771     & 9.419& 6.328    & 6.517 &8.3504 &7.862       \\              
$S/(B-\overline{B})$                               &362.679    & 369.857 &236.14 & 252.001 &280.8769  &284.407 \\          
$(h^++h^-+h^0)/(B-\overline{B})$                   &71.265     & 70.996 &53.551  & 54.479 & 62.108  &60.678 \\                  
$(h^++h^-+h^0)/(h^++h^-)$                          &1.6427     &1.64  &1.6338    & 1.638 &1.6414  & 1.641  \\                  
$(h^+-h^-)/(h^++h^-)$                              &0.0142     & 0.0143 &0.0184 & 0.0181 &0.0167  &  0.0169 \\                 
$E/(h^++h^-)$                                      &0.858     & 0.882 &0.9689   &1.004  &0.952  &  0.986  \\                 
$(s+\overline{s})/(h^++h^-)$                       &0.225     &0.218  &0.193   &0.196  & 0.221 &   0.213 \\                
$S/(h^++h^-)$                                      &8.360     & 8.546 &7.205   &7.576  &7.423 & 7.696  \\                  
$E/h^-$                                            &1.748     &1.796  &1.982   &2.054  &1.943  &  2.01  \\                   
$(B-\overline{B})/h^-$                             &0.0469     & 0.0471 &0.0624  &0.0612 &0.0534  & 0.0552\\                   
$(s+\overline{s})/h^-$                             &0.4588     & 0.443 &0.395  &0.401  &0.451  &  0.435\\                 
$S/h^-$                                            &17.0297    & 17.417 &14.738 &15.493  &15.161 &   15.719\\    
$(h^++h^-+h^0)/h^-$                                &3.346     &  3.341 &3.342  &3.349  &3.352  &  3.354\\                  
$(B+\overline{B})/h^-$                             &0.256      & 0.256  &0.327 &0.323  &0.289 &  0.295 \\  
$S/(s+\overline{s})$                               &37.118     &39.266  &37.314 &38.669  &33.636  & 36.176 \\                 
$S/E$                                              &9.7443     &9.694  & 7.436 & 7.544 & 7.801  & 7.804 \\    
$\frac{(h^++h^-+h^0)_{weak}}{(h^++h^-+h^0)_{total}}$  &0.325    &0.326  & 0.425 & 0.427  & 0.410  & 0.404 \\                 
\hline
\end{tabular}
 \end{table*}
%%%%%%%%%%%%%%%%%%%%%%%%%%%%%%%%%%%%%%%%%%%%%%%%%%%%%%%%%%%%%%%%%%%%%%%%%%%%%%%%%%%%%%%%%%

In conclusion, using the SHARE package 
we have analyzed the Au--Au $\sqrt{s_{NN}}=130$ GeV experimental output
 with a variety of statistical models.
All are able to fit particle ratios and total charged multiplicity, 
achieving the highest statistical significance
with the full non-equilibrium ansatz.
We have calculated bulk thermodynamic properties 
of the system from the fitted parameters, and found that
many of these are model dependent, with particularly
 strong discrepancies between full non-equilibrium ($\gamma_{q,s} \ne 1$)
and the rest.
We hope that higher statistics and 200 GeV data \cite{future}
 will be able to falsify some of these models unambiguously, and hence
allow a precise determination of the statistical properties of the system.

%%%%%%%%%%%%%%%%%%%%%%%%%%%%%%%%%%%%%%%%%%%%%%%%
\subsection*{Acknowledgments}
%\vspace*{.5cm}
Work supported in part by grants from: the U.S. Department of
Energy  DE-FG03-95ER40937 and DE-FG02-04ER41318,  NATO Science Program PST.CLG.979634,
the Natural Sciences and Engineering Research
Council of Canada,  the Fonds Nature et Technologies of Quebec.\\
%LPTHE, Univ.\,Paris 6 et 7 is: Unit\'e mixte de Recherche du CNRS, UMR7589.\\
G. Torrieri wishes to thank the organizers of the ``Focus on Multiplicity''
 conference (Bari, 2004),
 and the Tomlinson Foundation of McGill University for their generous  support.

\section*{References}
%%%%%%%%%%%%%%%%%%%%%%%%%%%%%%%%%%%%%%%%%%%

\end{document}